\documentclass{pasa}%

\usepackage{graphicx}

\title{Simulations of ionospheric refraction on radio interferometric data}

\author[Chege et al.]{J. Kariuki Chege$^{1,2}$\thanks{Corresponding author: J. K. Chege \newline \href{jameskariuki31@gmail.com}{jameskariuki31@gmail.com} }, C. H. Jordan$^{1,2}$, C. Lynch$^{1,2}$, J. L. B. Line $^{1,2}$ and C. M. Trott$^{1,2}$
\affil{$^{1}$International Centre for Radio Astronomy Research, Curtin University, Bentley, WA 6102, Australia}%
\affil{$^2$ARC Centre of Excellence for All Sky Astrophysics in 3 Dimensions (ASTRO 3D), Bentley, Australia}
}%

\jid{PASA}
\doi{10.1017/pas.\the\year.xxx}
\jyear{\the\year}

\usepackage{aas_macros}
\usepackage{hyperref}
\usepackage{placeins}
\usepackage{subfig}
\usepackage{multirow}
\hypersetup{colorlinks,citecolor=blue,linkcolor=blue,urlcolor=blue}

\begin{document}

\begin{frontmatter}
\maketitle

\begin{abstract}
The Epoch of Reionisation (EoR) is the period within which the neutral universe transitioned to an ionised one. This period remains unobserved using low-frequency radio interferometers which target the 21 cm signal of neutral hydrogen emitted in this era. The Murchison Widefield Array (MWA) radio telescope was built with the detection of this signal as one of its major science goals. One of the most significant challenges towards a successful detection is that of calibration, especially in the presence of the Earth's ionosphere. By introducing refractive source shifts, distorting source shapes and scintillating flux densities, the ionosphere is a major nuisance in low-frequency radio astronomy. We introduce \textsc{sivio}, a software tool developed for simulating observations of the MWA through different ionospheric conditions estimated using thin screen approximation models and propagated into the visibilities. This enables us to directly assess the impact of the ionosphere on observed EoR data and the resulting power spectra. We show that the simulated data captures the dispersive behaviour of ionospheric effects. We show that the spatial structure of the simulated ionospheric media is accurately reconstructed either from the resultant source positional offsets or from parameters evaluated during the data calibration procedure. In turn, this will inform on the best strategies of identifying and efficiently eliminating ionospheric contamination in EoR data moving into the Square Kilometre Array era.
\end{abstract}

\end{frontmatter}

\section{INTRODUCTION }
\label{sec:intro}
One of the principal science goals for low-frequency radio interferometers is to observe the era referred to as the Epoch of Reionisation (EoR) through the redshifted 21 cm signal. The 21 cm signal is emitted by neutral hydrogen, which was ubiquitous in the universe before the formation of the first luminous sources. As the first stars and galaxies began forming, their radiation gradually reionised the neutral hydrogen. Observing the spatial and spectral properties of the 21 cm signal can provide a valuable probe for the EoR, shedding light on the properties of these first galaxies (see reviews in, \citealt{Fan2006, Furlanetto2006, Pritchard2012, Zaroubi2013}). Low-frequency arrays attempting to observe the 21 cm signal include the Murchison Widefield Array (MWA, \citealt{Bowman2013, Tingay2013}), Long Frequency Array (LOFAR, \citealt{VanHaarlem2013}) and the Hydrogen Epoch of Reionisation Array (HERA, \citealt{Deboer2017}). Before a successful detection can be made, however, several challenging problems must be overcome, including the Earth's ionosphere.

\subsection{The ionosphere}
The ionosphere is a key challenge to not only EoR radio observations but to the calibration of all low-frequency radio observations in general \citep{Marti-Vidal2010, Tasse2013, Prasad2014, jordan2017, Trott2018}. In this work, we focus on the EoR science case, as it demands high levels of calibration accuracy, and (at a high level) cannot tolerate the ionosphere as a contaminant. This required calibration precision is due to the faintness of the 21 cm signal as compared to the much brighter emissions mainly from galactic synchrotron emission and extragalactic sources, usually referred to as the EoR foregrounds. 
 
The ionosphere is composed of layers of ionised material that behaves like a plasma between altitudes of  100 to 1000 km. Ionospheric structure and activity vary based on many factors such as Earth location, time, altitude, solar cycle, and in recent years the structure has been vigorously studied (e.g. \citealt{Davies1990}).  Solar activity is, however, the main driver of ionospheric activity with solar ultra-violet and extreme ultra-violet radiation increasing the ionisation during the day while recombination occurs during night hours. This results in temporal variations in the order of minutes or less with higher activity levels in the daytime as compared to night hours. The ionosphere is also permeated by the Earth's magnetic field and extreme geomagnetic activity can also result in dynamic ionospheric structures \citep{Loi2015a} The free electron column density of the ionosphere is usually referred to as the total electron content (TEC) and is measured in TEC units ($1$ TECU $=10^{16}$ m$^{-2}$).
 
Incident radio waves from distant galaxies undergo various propagation effects as they traverse the ionosphere (see e.g. \citealt{Thompson2017a}). These effects are due to a varying refractive index caused by spatial inhomogeneity of both the free electron column density and the magnetic field strength along the propagation path. Assuming that the magnetised plasma of the ionosphere is cold and collisionless, the refractive index can be estimated by a generalised Appleton-Hartree equation \citep{Shkarofsky1961}. This equation also assumes that the wave has a frequency $\nu$ much higher than the plasma frequency $\nu_p$. In turn, the Appleton-Hartree equation can be expanded into a Taylor series estimation (e.g. \citealt{DeGasperin2018}). The second-order term of this expansion is the most dominant and is the focus of this work. The third-order term applies for polarized signals in the form of Faraday rotation. In this work, we do not consider polarisation effects or any more of the higher-order terms.

The phase change ($\Theta$) due to the excess path length introduced by the varying refractive index in the line of sight ($l$) is described by:

\begin{equation}
\label{phase_delay}
    \Theta = -\frac{q_e^2}{4 \pi \epsilon_{\circ}m_ec\nu} \int{n_e dl},
\end{equation}

where $q_e$ is the electron charge, $\epsilon_o$ is the permittivity in free space, $m_e$ is the electron mass, $c$ is the speed of light, $n_e$ is the free electron density and $\nu$ is the frequency. The phase delays in Equation \ref{phase_delay} result in vector shifts in the expected positions of observed sky sources over the field of view.
The source offsets in radians can be described by equation \ref{phase_offsets}:

\begin{equation}
\label{phase_offsets}
    \Delta\theta = \frac{1}{8\pi^2} \frac{e^2}{\epsilon_o m_e}  \frac{1}{\nu^2}  \Delta STEC \; ,
\end{equation}
where STEC is the slant TEC of the ionosphere, applicable when the line of sight is at an arbitrary azimuth and elevation. We have assumed that the scattered amplitude of the propagating wave is much smaller compared to the incident amplitude. This is the weak scattering regime wherein we can neglect the higher spatial derivatives of STEC which cause interference and diffraction effects. Equations \ref{phase_delay} and \ref{phase_offsets} show that these ionospheric effects are dependent on the observing wavelength and therefore, lower frequency observations are more severely affected.

The spatial characteristic scales of an ionospheric medium can be described by the diffractive scale ($r_{\text{diff}}$, the scale at which the phase variance of the turbulence reaches $1$ rad$^2$), and estimated by use of the phase structure function given by:
\begin{equation}
\label{struc_func}
    D(r) = \left<(\phi(r_{\circ})-\phi(r_{\circ}+r))^2\right> \approx \left(\frac{r}{r_{\text{diff}}}\right)^{5/3},
\end{equation}
where the angle brackets indicate an ensemble average.
The $5/3$ index is typical for media following pure Kolmogorov turbulence as is mostly the case with the ionosphere \citep{Kolmogorov1941, Kolmogorov1991, Rufenach1972, mevius2016, jordan2017}.

\cite{lonsdale2005} identified 4 observational regimes within which current radio telescopes can be categorised in relation to the ionosphere. They are mainly based on three factors, namely:
\begin{enumerate}
    \item The characteristic scales of the ionospheric activity,
    \item The field of view (FoV) of the array,
    \item The physical size of the array determined by the length of its baselines.
\end{enumerate}

Arrays with fields of view smaller than the characteristic ionospheric structures can have all antennas observing through a linear ionospheric TEC gradient. Furthermore, if the baseline lengths are shorter than the ionospheric scales, the phase difference is negligible across all baselines and the ionosphere poses no problem to interferometric data in this case. Alternatively, if the baselines are longer, the overall ionospheric effect is direction independent and its correction during calibration can be a straight forward overall phase shift in the visibilities. The above are regimes 1 and 2 respectively.

The MWA, having a large FoV ($\sim 610$ deg$^2$ at $150$ MHz) and baselines lengths that are on the order of ionospheric scales, falls in regime 3. This results in DD ionospheric effects. The observed phase difference for each direction and baseline is proportional to the baseline's length as projected onto the ionospheric screen. Lonsdale's regime 4 applies to arrays with large FoV and baselines longer than those of the MWA, such as in arrays with different stations e.g. LOFAR. Here each station views a different ionosphere and can be calibrated separately. As described above, these regimes especially 3 and 4 with DD ionospheric effects act as good determinants on the calibration procedure applicable to the observed data \citep{Smirnov2011a}.

\subsection{Probing the ionosphere from its effects on  radio data}
\label{ionospheric_characterisation}
Various calibration schemes that target individual directions towards sources or patches in the sky have been developed over the past few decades. This is in the so-called direction dependent (DD) effects calibration regime. This progress has made it possible for interferometers to observe at low-frequencies with most of the data affected by the ionosphere being correctable. In the case of EoR science using the MWA data, the Real-Time System (RTS, \cite{Mitchell2008}, see Section \ref{test2}) is used to correct for ionospheric effects. 

Consequently, these calibration algorithms have also been used as tools to probe the ionosphere itself from its manifestation in the visibilities. Much effort has been made to reconstruct structures in the ionosphere from calibration data \citep{Loi2015, mevius2016, jordan2017, Helmboldt2020, Albert2020a, Albert2020b}. \cite{jordan2017} used ionospheric offsets toward each of the sky-model sources produced by the RTS during calibration to devise a metric system that gives the level of ionospheric activity at the time of the observation as a single value quality assurance metric (QAM). They also categorized ionospheric activity based on the characteristic scales and dominant directions of the offsets. 

Recently, \cite{Albert2020a} introduced Kriging to perform 3D tomographic inference of the ionosphere. They introduce a regression kernel that provides more accurate modelling than other kernels usually used in the Gaussian process regression field. In \cite{Albert2020b}, ionospheric screens inferred using this approach have been applied in addition to the facet-based DD calibration of LOFAR data resulting in less ionospheric-induced image plane artefacts. Other related works include \citealt{Loi2015} and \citealt{Helmboldt2020} mentioned later in this work, and \cite{Arora2015} who attempted to obtain ionospheric TEC gradients over the MWA using GPS measurements.

\subsection{\textsc{SIVIO}}
Our work is motivated by characterising the impact of the ionosphere on EoR science as well as devising ways of avoiding and/or mitigating ionospheric effects. At the time of writing, data suspected to be significantly corrupted by the ionosphere are usually omitted from further analysis by the MWA EoR team (e.g. \citealt{Trott2020}). This is, for instance, datasets with $r_{\text{diff}} \ll 5$ km at $\sim150$ MHz which typically shows high ionospheric contamination and is usually not used for EoR science analysis \citep{mevius2016, vedantham2016, jordan2017}. Similar challenges occur to most MWA science cases including for example the GLEAM survey \citep{Wayth2015} observations where a part of the sky coverage was removed due to ionospheric effects. There remains, in every survey, a significant percentage of the data that is deemed irreparably affected by the ionosphere.
Thus, there is a need for a method to isolate ionospherically affected visibilities in order to quantify the impact of the ionosphere on specific science cases.

We introduce the IOnospherically contaminated Visibilities SImulator (\textsc{sivio} \footnote{Acronym obtained from capitalized letters read in reverse.}). This is a tool that simulates realistic radio interferometric mock observations in the presence of user-defined ionospheric conditions. Different synthetic ionospheric structures as observed by prior MWA surveys can be generated and the contaminated output visibilities can be analysed. This helps to explicitly seclude ionospheric contamination effects from other observational effects and study their manifestations in specific science cases such as the EoR project.

\textsc{sivio} is an extension to the work done by \cite{jordan2017}. The \textsc{cthulhu}\footnote{\hyperlink{https://gitlab.com/chjordan/cthulhu}{https://gitlab.com/chjordan/cthulhu}} software suite developed in that work reconstructs TEC gradients from a collection of right ascension and declination offsets such as those obtainable from the RTS calibration outputs (see Section \ref{test3}). \textsc{sivio} complements \textsc{cthulhu} by making it possible to control the level of ionospheric activity and produce visibilities that can be calibrated and analysed in both the visibilities as well as the image plane. The simulations can also be used to examine the direction-dependent performance of a calibration scheme like the RTS (see Section \ref{test3}). The method used is described in Section \ref{sec:methods}, while the tests and results are shown in Section \ref{sec:tests}. Conclusions and various possible future use cases are discussed in Section \ref{sec:discuss}. \textsc{sivio} code can be accessed publicly on \textsc{github}\footnote{\hyperlink{https://github.com/kariukic/sivio}{https://github.com/kariukic/sivio}} and a documentation\footnote{\hyperlink{https://sivio.readthedocs.io/en/latest/}{https://sivio.readthedocs.io/en/latest/}} with the installation and usage guidance is also available.

\section{Method}
\label{sec:methods}
\begin{figure}
    \centering
    \includegraphics[width=8cm]{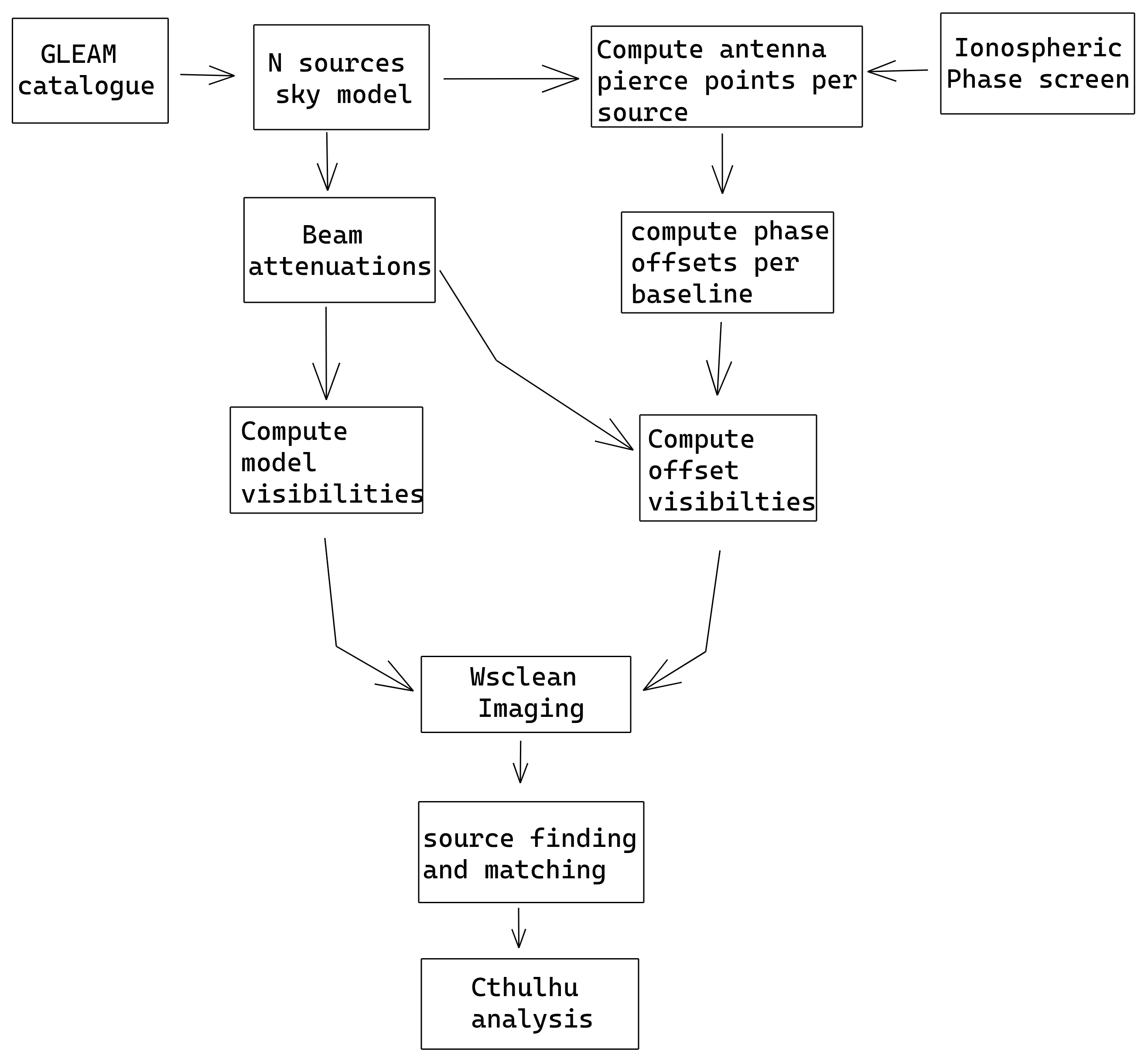}
    \caption{\textsc{Sivio} pipeline: \textsc{Sivio} simulates both the model and ionospherically contaminated visibilities. It also provides functionalities to analyse the simulated ionospheric effects in the image plane.}
    \label{fig:sivio pipeline}
\end{figure}

The input to \textsc{sivio} is a \textsc{CASA}\footnote{\hyperlink{https://casa.nrao.edu/casadocs/casa-5.1.0/reference-material/measurement-set}{https://casa.nrao.edu/casadocs/casa-5.1.0/reference-material/measurement-set}} measurement set \citep{Mcmullin2007}, which is a standard table system widely used to store radio  interferometric data. The input measurement set provides the observation parameters including the array antenna configuration, the observation time, the phase center and the sampling of the $(u,v)$ plane (See Section \ref{subsec:truesky}). The measurement set is also used to store the output simulated visibilities. \textsc{sivio} offers standalone true sky visibility simulation functionality for comparison with the contaminated ones produced. Furthermore, \textsc{sivio} provides functionality for performing source extraction using the \textsc{aegean} source finder \citep{Hancock2012} as well as source matching between the model sky and the contaminated sky images. The structure of \textsc{sivio} algorithm is shown in Figure \ref{fig:sivio pipeline}. The following Subsections describe in detail the workings of \textsc{sivio}.

\subsection{True sky visibilities}
\label{subsec:truesky}
The algorithm begins by creating a sky model, $I$, obtained from a sky source catalogue that can be generated in either of two ways. One could use a uniform spatial distribution of point sources with flux densities obtained using a broken power-law as applied in \cite{Franzen2016} and  \cite{Trott2018}. The alternative is by choosing N sources from the galactic and extra-galactic all-sky MWA survey (GLEAM) catalog \citep{Hurley-Walker2017}. These sources are selected to have a flux density exceeding a given minimum flux density and within a user-specified sky area around the observation pointing center. The flux densities of the sources for each frequency within the observing bandwidth are calculated using the spectral index provided by GLEAM.

\begin{figure}
    \centering
    \includegraphics[width=.5\textwidth]{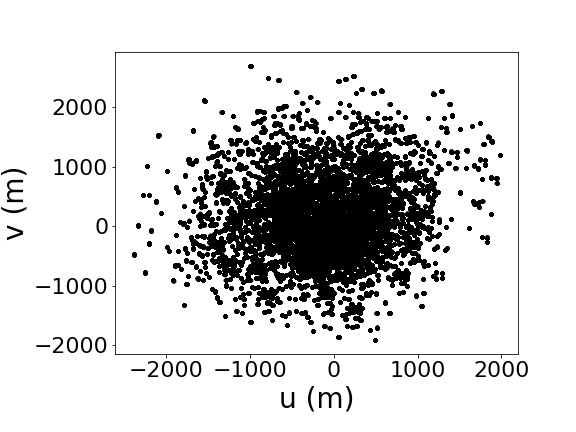}
    \caption{MWA phase 1 $(u,v)$ plane for a $\sim2$ minutes snapshot observation. Each baseline of a telescope array samples a single Fourier mode of the sky (black dots) and an image of the sky can be obtained by Fourier transforming the visibilities.}
    \label{uvplane}
\end{figure}

Subsequently, we obtain the apparent flux densities as would be observed by the MWA telescope by applying the instrumental flux density attenuation per source direction, derived from the beam model described in \cite{Sokolowski2017}. These apparent flux densities values are the ones used in the visibility measurement equation which is given by \citep{Thompson2017}:

\begin{equation}\label{ms_eqn}
\begin{split}
V(u,v,w)& =\int\int{A(l,m)I(l,m) \,\, \times} \\
 & {e^{-2\pi{i}(ul+vm+w(\sqrt{1-l^{2}-m^{2}}-1))}  \frac{dl \, dm}{\sqrt{1-l^{2}-m^{2}}}},
\end{split}
\end{equation}

where $(u, v, w)$ is a cartesian coordinate system used to describe the baselines, with $w$ pointing to the direction of the phase centre, $A(l,m)$ represents the instrumental response as a function of the effective collecting area and the direction of the incoming signal and $(l, m)$ are the directions cosines. Figure \ref{uvplane} shows a sample MWA phase 1 configuration $(u,v)$ plane that would be used in equation \ref{ms_eqn} to calculate the visibilities for a $\sim2$ minutes snapshot observation.

Radiometric measurement noise is then added to the visibilities as:
\begin{equation}
    \label{thermal}
    \sigma_N = 10^{26}\frac{2k_\text{B} T_{\text{sys}}}{A_{\text{eff}}}\frac{1}{\sqrt{\Delta\nu\Delta t}} \;\;\; \text{Jy}, 
\end{equation}
where $T_{\text{sys}}$ is the system temperature, $A_{\text{eff}}$ is the effective collecting area of the antenna, $\Delta\nu$ is the frequency resolution, and $\Delta t$ is the integration time of the visibilities. Typical values for the MWA are $T_{\text{sys}} = 240$ K, $A_{\text{eff}}=21$ m$^2$, $\Delta\nu = 40$ KHz and $\Delta t = 8$ s. These data are written back into the DATA column of the measurement set.

\subsection{Ionospherically contaminated visibilities}
\label{subsec:ionosky}
The ionospherically contaminated visibilities are the primary output of \textsc{sivio}. To capture the ionospherically-induced phase shifts observed by an interferometer into the visibilities, we apply the differential phase gradients representative of $\Delta STEC$, obtained from the simulated phase screens to equation \ref{ms_eqn}. The steps involved are as follows:
\begin{enumerate}
    \item Constructing a TEC screen.
    \item Obtaining pierce point coordinates on the TEC screen for each source and antenna.
    \item Evaluating the differential phase offsets per source per baseline.
    \item Computing the corrupted visibilities.
\end{enumerate}

\subsubsection{Phase screens and extraction of pierce points}
\label{subsec:phscreens}
We model the phase screen as a thin screen approximation of the ionosphere, essentially a plane assumed to be at a certain altitude from the ground plane of the array. This model has been found to be a sufficient approximation of the ionosphere and has been deemed suitable in, for instance, \cite{Intema2009, Marti-Vidal2010}. 

\cite{Koopmans2010} showed that the 3D ionosphere results in a spatially varying ionospheric scattering point spread function (PSF). At present, SIVIO has an invariant PSF as it models the ionosphere with a thin screen model, resulting in inaccuracies. \cite{Intema2009a} estimated the error resulting from using a 2-dimensional (2D) screen to represent the 3-dimensional (3D) ionosphere to be in the order of fewer than 10 degrees for up to 60 degrees zenith angles, and even less for shorter baselines at 74 MHz. Their comparison between a single 2D screen and multiple ones realised marginal improvement in real data calibration. We, therefore, expect the thin screen model to be reasonably accurate in the first 3 Lonsdales regimes and even more so for zenith-pointed observations. For the MWA, the phase I configuration should be the least affected by any thin-screen associated errors because of its compactness (112 of the total 128 antennas are within 750 m from the array centre).  In Lonsdale regime 4, a thin screen ionospheric model might be inadequate and the tomographic inference by \cite{Albert2020a} seems more accurate. However, we note that \cite{Koopmans2010} shows the importance of long baselines in correcting for the 3D w-term of the ionosphere as they can probe the largest ionospheric structure scales. This tentatively suggests that a thin screen might also be usable for the extended configuration in regime 4 but we do not explore that in this work.

\cite{Helmboldt2020} reconstructed ionospheric disturbances over many hours using data observed for the GLEAM survey \citep{Wayth2015}. Using these observations, the authors detail four different forms of disturbances. The first occurrence was of calm coherent anisotropic ionospheric structures which cover $\sim50\%$ of the total ionospheric conditions identified. The structures are in the orders of few milli-TECU km$^{-1}$ and spatial scales of 10s of degrees (hundreds of km). The others are, (1) field aligned structures in the top-side ionosphere similar to ones originally observed by \cite{Loi2015}. (2) East-West aligned structures which were linked to wind excess gravity waves, and (3) Northeast-to-Southwest structures linked to electrobuoyancy.

\textsc{sivio} provides phase screens analogous to the ones observed in real data with 2 inbuilt phase screens namely Kolmogorov-like ($k$), and TEC with duct-like quasi-isotropic structures ($s$). A phase screen is constructed as a rectangular plane with dimensions proportional to the sky area being observed, the height of the screen from the array's ground plane and the chosen scaling for the pixel size. The level of turbulence is a hyper-parameter that can be modulated as desired by the user. Alternatively, a user can choose to input their custom phase screen whose required length ($l$) and width ($w$) dimensions can be calculated by \textsc{sivio}, given the desired grid scaling, screen height, and FoV.

\begin{figure}
    \centering
    \includegraphics[width=.5\textwidth]{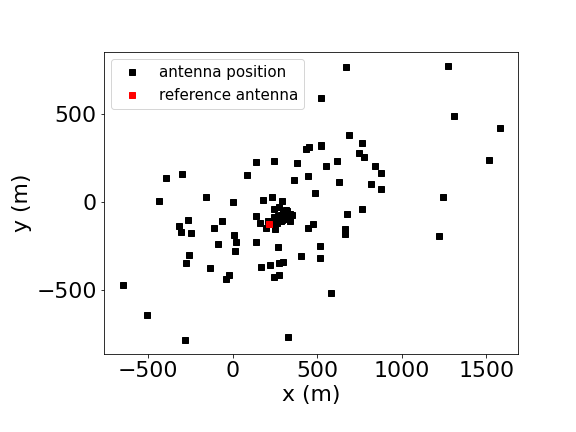}
    \caption{MWA phase I antennas positions with the reference antenna used to calculate the baseline vector offsets for each antenna shown as a red square.}
    \label{xy}
\end{figure}

To calculate the optimum phase screen dimensions for the observation, we first use the $(x,y,z)$ coordinates of individual antennas corresponding to the East, North, and up directions respectively from the input measurement set. Figure \ref{xy} shows the antenna positions of the MWA phase I with the chosen reference antenna used to calculate the baselines vector offsets.

Using the scaled maximum baseline vector, $d_{max}$, screen height $h$, and the maximum zenith ($\zeta'$) and azimuth ($a'$) angles observable in the field of view, the length ($q$) and width ($r$) needed for the screen are given by:

\begin{equation}
\label{screensize}
    q = 2(d_{max} + h\tan{\zeta'}\sin{a'}), \:\:\:\:\:\:    r = 2(d_{max} + h\tan{\zeta'}\cos{a'}).
\end{equation}

\begin{figure*}[!t]
    \centering
    \includegraphics[width=\textwidth]{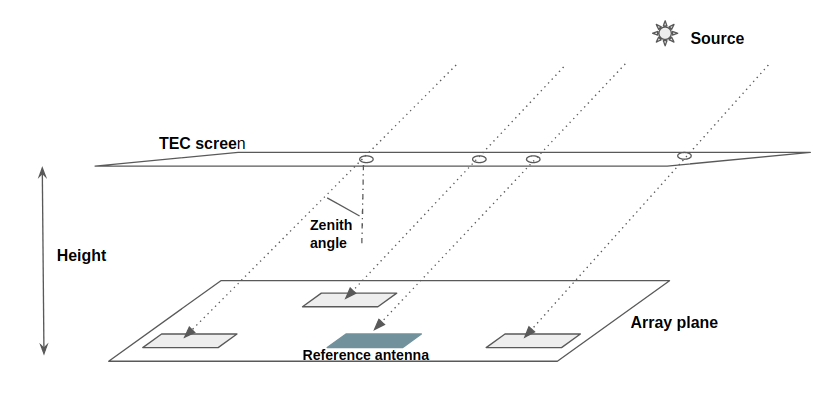}
    \caption{A schematic overview of a thin TEC screen at a given height above the array plane. The reference antenna is used as a point of origin for the array plane and directly above it is the TEC screen origin point. The zenith angle of the source is used to calculate the displacement of the pierce point from the TEC position vertically corresponding to the observing antenna.}
    \label{geometry}
\end{figure*}

The center of the phase screen is then aligned with the phase center of the observation and also to the reference antenna. With the scaled array plane and the TEC plane in place, we can now proceed to evaluate the pierce point of each antenna, based on its line of sight towards the source direction. Each source position $(\alpha, \delta)$ in the sky model is used to obtain the $n_{\text{antennas}}$ pierce points on the phase screen resulting into a total of $n_{\text{sources}} \times n_{\text{antennas}}$ pierce points. Figure \ref{geometry} is a toy representation of the simulation setup for a single source.

For a source at zenith angle ($\zeta$) and azimuth ($a$) its pierce point coordinates $(p_x, p_y)$ as observed from an antenna with $d$ distance scaled separation from the reference antenna can be obtained by:

$$(p_x, p_y) = ((d + h\tan{\zeta}\sin{a}), (d + h\tan{\zeta}\cos{a}))$$

\begin{figure*}[!t]
    \centering
    \includegraphics[width=\textwidth]{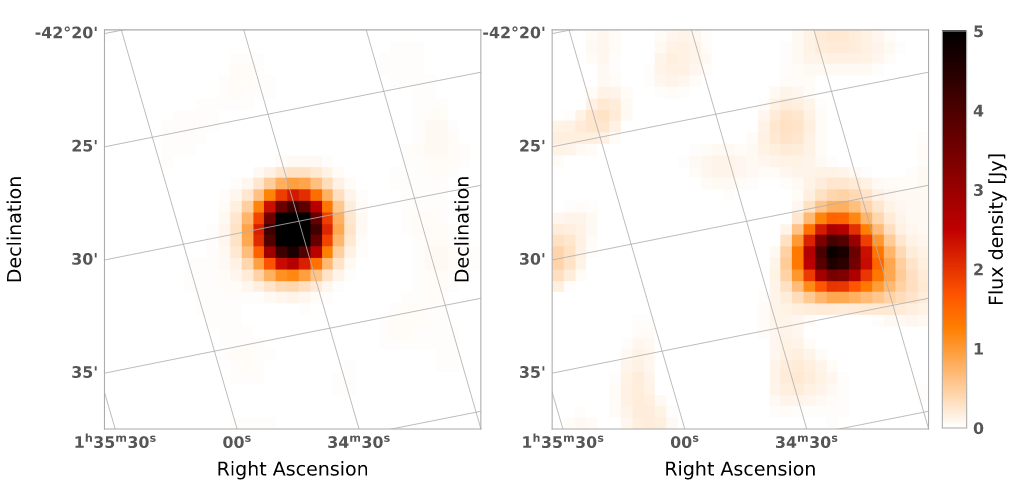}
    \caption{Simulated source positional offset in the image domain. Left: source at real position. Right: source shifted in position for both RA and Dec after applying a phase offset to each antenna}
    \label{posoffset}
\end{figure*}

The gradient value for each pierce point in the orthogonal $x$ and $y$ directions is then used as the additional phase offsets in the measurement equation when computing the visibilities of that source. A positional shift of the sources in the image domain can be implemented by injecting an additional phase term, $\phi'$, to the phase angle $\phi$ when computing the visibilities measurement equation (see figure \ref{posoffset}).

$$V(u,v,w)=\int\int{I(l,m)e^{-2\pi{i}(\phi+\phi')}}dl \, dm,$$
where
$$\phi = ul+vm+w(\sqrt{1-l^{2}-m^{2}}-1),$$ and,

$$\phi' = \Theta(p_{x_1}, p_{y_1}) - \Theta(p_{x_2}, p_{y_2})$$
is the difference between the additional phase shift, $\Theta$ underwent by each of the two parallel waves traversing the ionosphere and recorded by two antennas which form each baseline. Noise is similarly added as in Section \ref{subsec:truesky}.

\section{Verification}
\label{sec:tests}
In testing \textsc{sivio},  we aim to verify that antenna path delays are correct for each source and antenna, resulting in accurately contaminated visibilities. We conduct 3 experiments as shown in Sections \ref{test1}, \ref{test2} and \ref{test3}. An example of a full \textsc{sivio} experiment may consist of the following steps which can all be done conveniently using \textsc{sivio}:

\begin{enumerate}
    \item Obtaining an input measurement set.
    \item Choosing the sky area size for which to obtain a sky model source catalog from.
    \item Constructing the input phase screen (either $k$ or $s$) with desired turbulence level.
    \item Simulating the true and contaminated visibilities.
    \item Imaging, source extraction and source matching from the two catalogues.
    \item Calculating source offset vectors in RA and Dec from the true positions.
    \item Reconstructing the output phase screen.
\end{enumerate}

\subsection{Test 1: Ionospheric offsets frequency dependence.}
\label{test1}
\begin{figure}[t]
    % \centering
    \includegraphics[width=.5\textwidth]{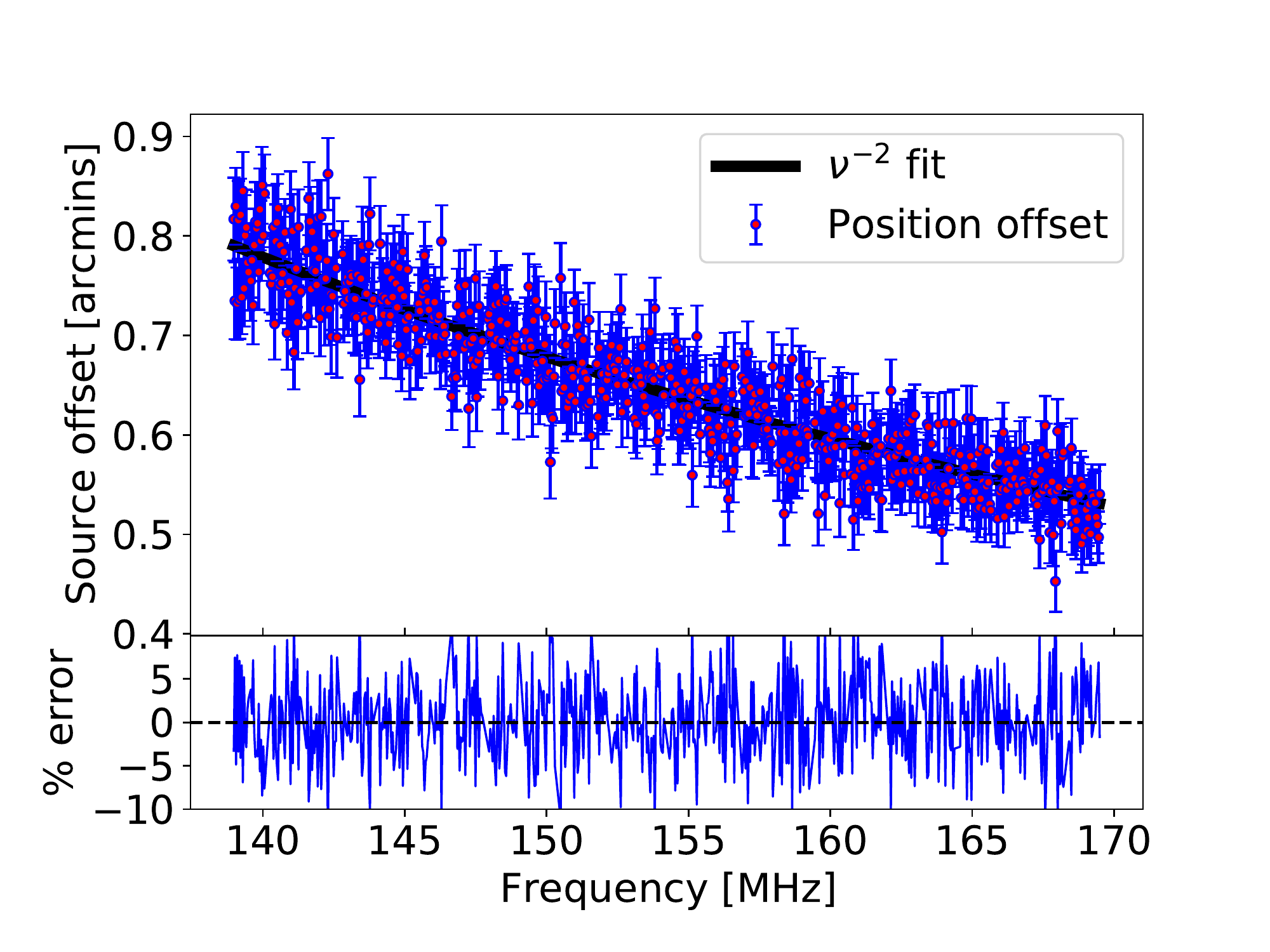}
    \caption{Top panel: Source position shifts as a function of frequency fitted with an equation \ref{phase_offsets} model (black line). The error bars are 2 sigma uncertainties propagated from the position error given by the source finder. Bottom panel: Percentage error per frequency between the offsets and the model.}
    \label{fig:channel_shifts}
\end{figure}

The first test aims to verify that the positional offsets introduced into the simulated visibilities follow the $\lambda^2$ relation from equation \ref{phase_offsets}. For this, we simulate a single bright source at zenith. This is where, for a zenith-pointing observation,  the flux attenuation by the instrumental response is minimum. From both sets of visibilities (the true and ionospherically contaminated), we image each individual channel, run a source finder, then calculate the position shift of the source. Figure \ref{fig:channel_shifts} shows the shift in the source position as a function of frequency with an equation \ref{phase_offsets} fit. The error bars are 2 sigma ($\sigma_{\alpha\delta}$) uncertainties propagated from the position error  evaluated by the source finder as shown in Equation \ref{eqn:errorsprop}. 

\begin{equation}
    \label{eqn:errorsprop}
    \sigma_{\alpha\delta} = \frac{1}{2}\text{Dist}\left(\frac{2\frac{\Delta\alpha}{|\alpha-\alpha_{o}|}{\alpha}^2+2\frac{\Delta \delta}{|\delta-\delta_{o}|}{\delta}^2}{{\alpha}^2+{\delta}^2}\right)
\end{equation}

where $\text{Dist}$ is the spherical distance between the true, ($\alpha$, $\delta$), and the offset, ($\alpha_{o}$, $\delta_{o}$), source positions respectively, and ($\Delta\alpha$, $\Delta\delta$) are the quadrature of the RA and Dec errors provided by the source finder for both the true and offset source positions. The percentage root mean square error between the model and the position offsets was found to be $\sim 0.3\%$ which shows that we adequately capture the frequency dependence of the ionospheric offsets.

\subsection{Test 2: Direct TEC reconstruction from uncalibrated visibilities}
\label{test2}
\begin{figure*}[t]
    \centering
    \includegraphics[width=\textwidth]{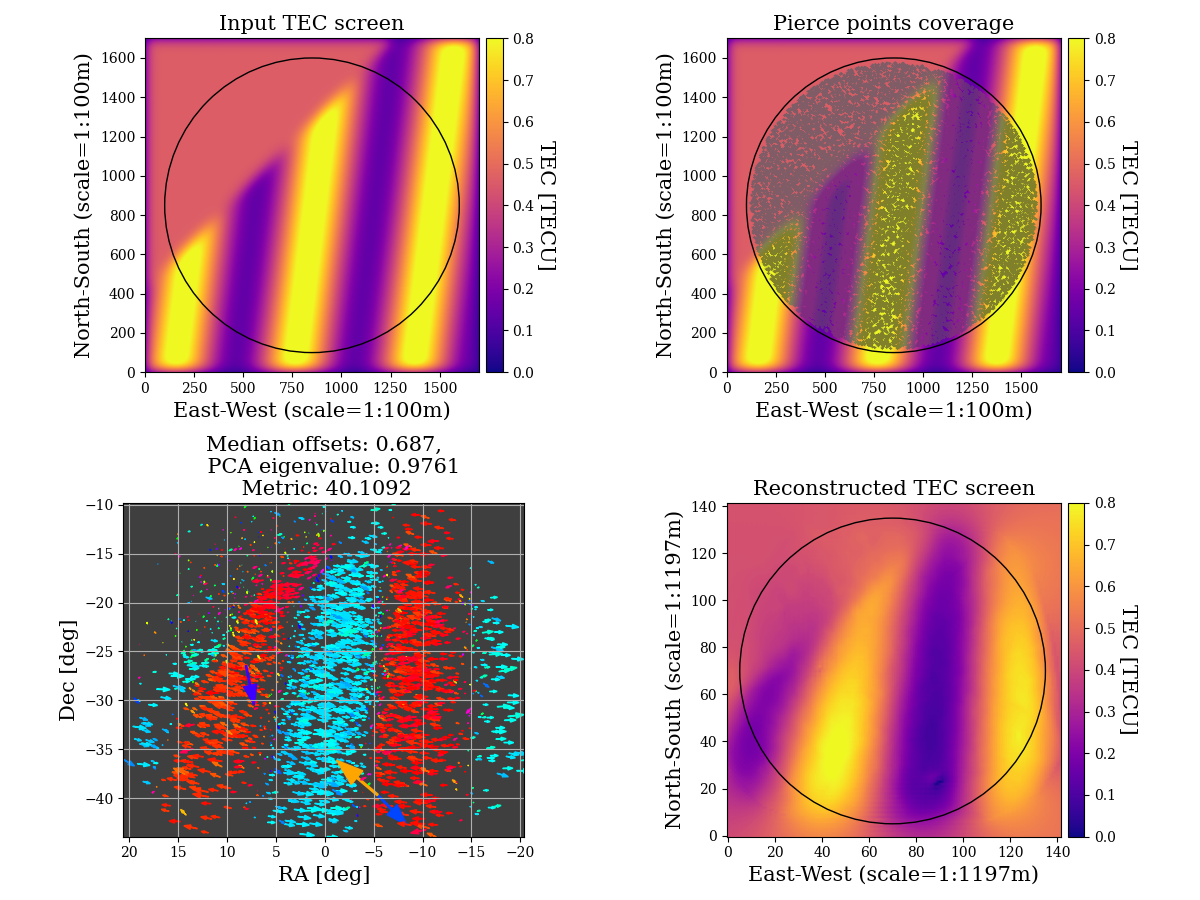}
    \caption{Top left: simulated $s$-screen. Top right: Phase screen overlaid with pierce points for GLEAM sources above 0.1 Jy in a 20 deg sky radius area centered at RA=$0.0^{\circ}$ and Dec$=-27.0^{\circ}$. Bottom left: Vector offsets for each source at 154 MHz. Bottom right: Reconstructed TEC from the vector offsets. The black circle specifies the reconstructed area of the input screen.}
    \label{fig:stec}
\end{figure*}

\begin{figure*}[t]
    \centering
    \includegraphics[width=\textwidth]{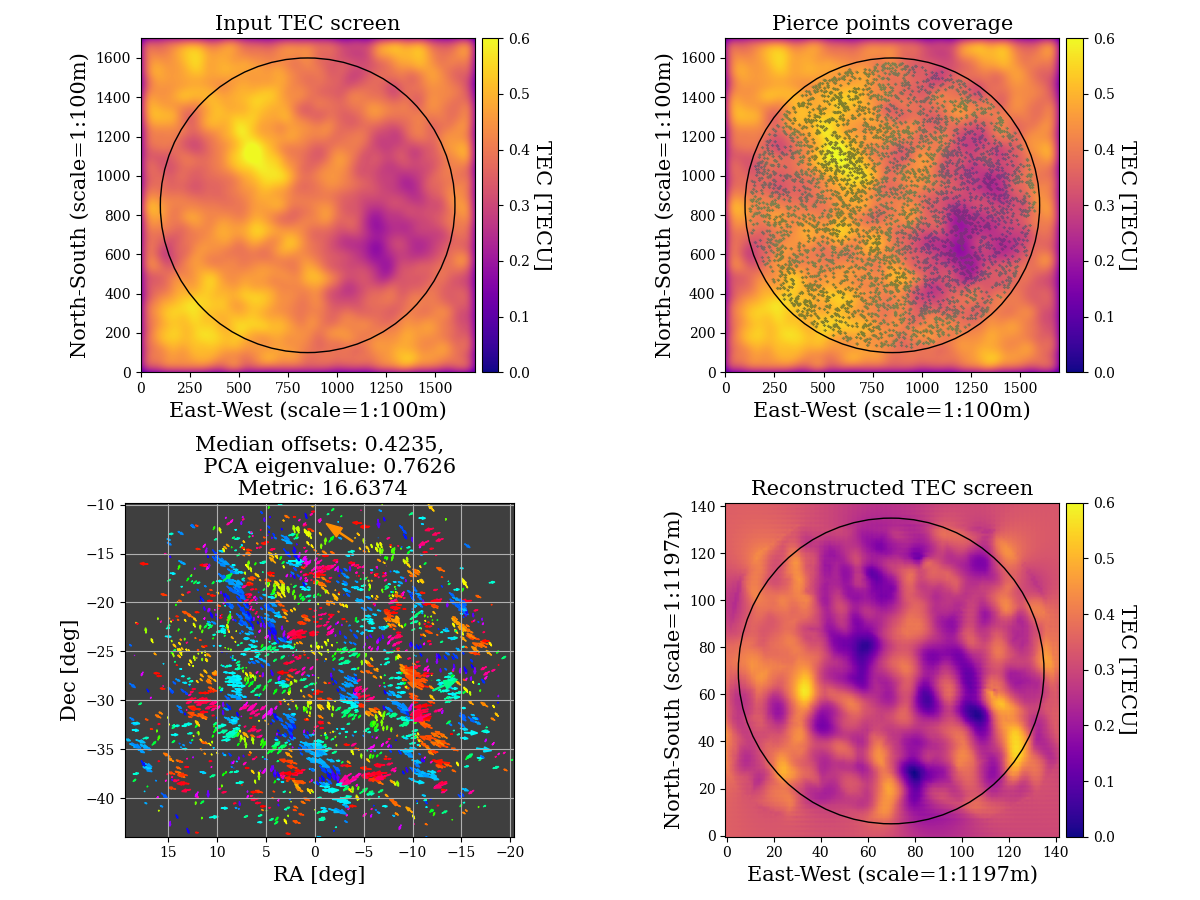}
    \caption{Top left: simulated $k$-screen. Top right: Phase screen overlaid with pierce points for GLEAM sources above 0.4 Jy in a 20 deg sky radius area centered at RA=$0.0^{\circ}$ and Dec$=-27.0^{\circ}$. Bottom left: Vector offsets for each source at 154 MHz. Bottom right: Reconstructed TEC from the vector offsets. The black circle specifies the reconstructed area of the input screen.}
    \label{fig:ktec}
\end{figure*}

By simulating contaminated visibilities dominated by ionospheric effects only, we should be able to image them directly and observe the source shifts of the sources without the need for prior calibration. It should be noted that this is however not usually the case with real data where at least a direction independent calibration run is mandatory before any directional ionospheric effects can be assessed. An experiment that incorporates full calibration as would be applied to real data is described in Section \ref{test3}.

For this test, we constructed a ducted ($s$-screen) and a Kolmogorov ($k$-screen) TEC phase screens both at $h=200$ km. For a 20 deg sky radius, this corresponds to a phase screen of $\approx 160$ km in relative latitude and longitude. For the $s$-screen we use a GLEAM sky model with a minimum flux density cutoff of $0.1$ Jy resulting in $\sim 30000$ sources. The number of pierce points required for sufficient sampling of an ionospheric TEC screen for the MWA can be as low as $\sim 500$. For this reason, in the $k$-screen simulation we apply a 0.4 Jy cutoff resulting in $\sim3000$ sources in the sky model. Sample \textsc{sivio} output plots are shown in Figures \ref{fig:stec} and \ref{fig:ktec}. We note that the differential TEC values used here are on the higher extreme and they do not represent typical levels.  In both figures, the top left plot shows the phase screen in degrees, pierce points are overlaid onto the phase screen on the top right plot. The bottom left plot shows the vector offsets for each source after source catalog matching, and the reconstructed TEC is shown on the bottom right. We see that the reconstructed TEC screen is qualitatively a relatively good match with the input screen. \textsc{cthulhu} applies a reconstruction of the offsets from sources estimated to be located within the primary beam causing a zoom-in effect. This might be the reason why the reconstruction of the Kolmogorov TEC screen is not visually as similar to the input as compared to the $s$ screen. However, the statistics obtained are very close to those of the input screen. A more quantitative analysis using statistics obtained from the \textsc{cthulhu} suite is given in Section \ref{test3}.

\subsection{Test 3: Calibration and TEC reconstruction from calibration products}
\label{test3}
\begin{figure*}[!ht]
    \subfloat{{\includegraphics[width=9cm]{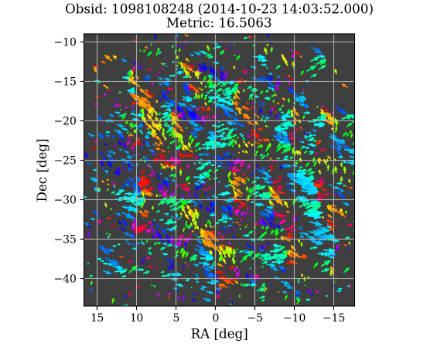} }}%
    \qquad
    \subfloat{{\includegraphics[width=9cm]{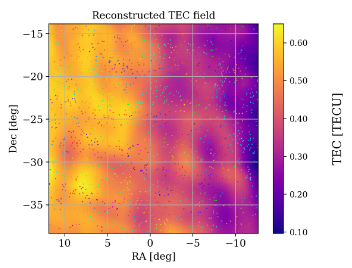} }}%
    \caption{Left: Measured vector offsets from the RTS calibration data products at 154 MHz and used to calculate the $Q$ metric of the simulation. Right: Reconstructed TEC from the offsets. The ionospheric effects simulated into the visibilities by \textsc{sivio} are accurately detected and quantified by the direction-dependent calibration of the RTS.}%
    \label{rtstec}%
\end{figure*}

As aforementioned, a possible use case for \textsc{sivio} can be as an assessment to a data calibration scheme and calibration is a vital step in radio data processing. Therefore, for completion, we want to see how a typical radio astronomy calibration regime such as the RTS would quantify the phase offsets introduced in our simulated data. For this, we run \textsc{sivio} with the same $k$ phase screen used in figure \ref{fig:ktec} and calibrate the corrupt visibilities through the RTS as would be done for EoR experiments with the MWA.

We first summarize the procedure applied by the RTS algorithm. A detailed account can be found in \cite{Mitchell2008}. The RTS Section relevant for ionospheric and instrumental gain correction is a peeling calibration iteration called the calibration measurement loop (CML). The total 30.72 MHz bandwidth of the visibilities is subdivided into 24 subbands of 1.28 MHz each of which is further split into 40 kHz width fine channels. This results in 768 fine channels for the whole bandwidth.  The CML mostly runs the calibration simultaneously for each fine channel with a few instances in the process where the whole bandwidth is used to fit for frequency-dependent gain parameters.

First, a pool of the apparently brightest sources (calibrators) based on the beam response in their sky direction is accumulated. The contribution of all these sources to the visibilities is then subtracted. This is in order to minimize sidelobes from other sources when peeling each calibrator individually. In their descending order of brightness, each calibrator is then re-added back into the visibilities and a rotation is performed shifting its estimated position to the phase center of the observation where the instrumental response is maximum. This makes the calibrator dominant in the visibilities. Since the source is expected to be exactly at the phase center, its $(l,m)$ values from equation \ref{ms_eqn} are expected to be zero. Deviations from this are attributed to ionospheric refraction and can be quantified enabling a $\lambda^2$ fit similar to the one in Section \ref{test1} to be carried out. A linear least-squares solution is then computed to obtain a model of the ionospheric refraction. These steps are run in iteration for the calibrator and if the fitted ionospheric gains satisfy given goodness of fit tests, they will be used to perform a peel of the calibrator. This process is carried out for all the calibrators in the pool.
With the gains derived from the RTS in this way, we can use \textsc{cthulhu} to obtain the offsets in both RA and Dec for each source in the visibilities and reconstruct the TEC screen. An example reconstructed TEC is shown in figure \ref{rtstec}.

The \textsc{cthulhu} suite evaluates the median position offset values, $m$, as well as the dominant principal component analysis (PCA) eigenvalue, $p$, of the offsets. These values are combined to produce a single quality assurance metric value, $Q$, given by,
\begin{equation}
\label{eqn:qam}
Q = \begin{cases} 
  25m+64p(p-0.6), & p> 0.6 \\
  25m, & \text{otherwise}
\end{cases}
\end{equation}
The quality metric can be interpreted as the level of ionospheric turbulence at the time of the observation. A higher $Q$ value can be as a result of either a large source offset, anisotropicity in the spatial ionospheric structure or both. \cite{Trott2018} showed that higher anisotropicity in the ionosphere degrades the resulting 21 cm power spectrum. For this reason, observations with $Q>5$ are usually not used in EoR power spectrum analysis. However, this threshold might vary for different science cases.

Table \ref{tab:1} summarizes the values obtained from \textsc{cthulthu} for both the test in Section \ref{test2} as well as the one in this Section. For the $k$ and $s$ phase screens in Section \ref{test2}, we obtain a quality metric of $Q=40.1$ and $Q=16.6$. After using the same $k$ phase screen and running it through the process described in this Section, we obtain a quality metric of $Q=16.5$. This value is equal to the $Q$ value obtained in Section \ref{test2} to an accuracy of $<1\%$. This agreement from the two tests further confirms that  \textsc{sivio} accurately contaminates the visibilities based on the provided phase screen. It also shows that these ionospheric effects can be perceived by a direction-dependent calibration algorithm like the RTS and, consequently, the contaminant ionosphere can be characterised as mentioned in Section \ref{ionospheric_characterisation}.

\begin{table}%[h]
\begin{tabular}{|l|l|l|l|l|}
\hline
Test               & Screen & Offset (m) & PCA (p) & Metric (Q) \\ \hline
\multirow{2}{*}{2} & s           & 0.6867            & 0.9762  & 40.1092       \\ \cline{2-5} 
                   & k           & 0.4235            & 0.7626  & 16.6374       \\ \hline
3                  & k           & 0.4965            & 0.7225  & 16.5063       \\ \hline
\end{tabular}
\caption{The median position offset ($m$ ,arcmins) and the dominant PCA eigenvalue ($p$) of the offsets for an $s$ (quasi-isotropic ducts) and $k$ (Kolmogorov turbulence) phase screen as run in Sections \ref{test2} and \ref{test3}. These values are combined to produces a single quality assurance metric value ($Q$) which can be used as a measure of overall ionospheric severity. The quality assurance metric from Section \ref{test3}, $Q = 16.5063$, is equal to the $Q$ value obtained in Section \ref{test2}, $Q = 16.6374$, to an accuracy of $<1\%$.}
\label{tab:1}
\end{table}

\section{Summary}
\label{sec:discuss}
We have introduced a software package that can be used to simulate radio observations and contaminate the visibilities with ionospheric effects obtained from simulated phase screens representing the ionosphere. The tests have shown that the offsets introduced to the visibilities conform to the expected $\lambda^2$ dependence. Furthermore, we have shown various methods that can be used to reconstruct the input phase screens.

Using the median offsets and the dominant PCA eigenvalue we have shown that various simulation outputs can be ranked on the level of ionospheric activity. We have also shown the capability of \textsc{sivio} to assess ionospheric calibration schemes such as the RTS buy using the ionospheric gains obtained during calibration to reconstruct the phase screen. The RTS accurately probes the input phase screen capturing the turbulence level as well as the spatial structures.

Position offsets modelled here are obtained from the first-order derivative of STEC and can be determined with frequency and STEC as the only parameters. Real ionospheric effects will however include the higher-order derivatives and these include effects such as visibility amplitude scintillation and source angular broadening (see e.g. \citealt{Loi2015a, DeGasperin2018}). Ionospheric absorption and emission can also cause slight variations in observed amplitudes (e.g. see \citealt{Sokolowski2015}). We do not attempt to model these in sivio as they are expected to be much less significant for MWA and can therefore be largely ignored \citep{Loi2015a}. However, they get more severe at frequencies below $\sim100$ MHz and are governed by many more parameters (see e.g. \citealt{vedantham2016, Vedantham2016a}). At a high level, these secondary effects are expected to be worse for longer baseline configurations and therefore the extent to which sivio can be used realistically at lower frequencies would be largely determined by the physical size of the array. For instance, the compact MWA configuration is advantageous because the higher-order effects are less apparent in its relatively short baselines. Additionally, due to its relatively lower angular resolution, fewer sources are resolved and the effect of scintillation is smaller than that of larger arrays.

Based on the simplifying assumptions applied, \textsc{sivio} simulations do not capture all physical aspects of a realistic ionosphere. Here we mention several shortcomings of sivio as currently implemented that would limit its usage in some science cases. First, it is worth mentioning that by assuming $\nu\gg\nu_p$, we are neglecting any ionospheric reflection effects. However, the plasma frequency can be multiple times higher than the usual $\sim10$ MHz, especially in low elevation directions and during the presence of high free electron density clouds (e.g. sporadic-E clouds). Polarimetry Studies are not applicable with \textsc{sivio} since we currently only consider total intensity visibilities. Additionally, we assume negligible ionospheric dispersion in the frequency bandwidth of the simulated observation. This might be an impediment in transients studies such as pulsar observations where precise dispersion measurements are required. Furthermore, the ionosphere is also not always in the weak scintillation regime; in rare events, it gets into the strong scintillation regime where even the usually negligible higher-order effects are magnified. We have implemented static ionospheric phase screens over the observation duration of the input measurement set. This might be a drawback for studying temporal effects of the ionosphere over longer timescales. However, such simulations with temporally varying screens can still be easily carried out using multiple measurement sets.

In future work, \textsc{sivio} simulations of different ionospheric conditions will be used to assess the tolerable levels of ionospheric activity in data used to calculate the EoR 21cm power spectrum using the MWA and inform on the optimal strategies for using the data.

\begin{acknowledgements}
Kariuki Chege thanks Torrance Hodgson, Ulrich Mbou Sob and Cyndie Russeeawon for useful radio interferometry and programming discussions during the early stages of this work. This work was supported by the Centre for All Sky Astrophysics in 3 Dimensions (ASTRO3D), an Australian Research Council Centre of Excellence, funded by grant CE170100013. CMT is supported by an ARC Future Fellowship through project number FT180100321. The International Centre for Radio Astronomy Research (ICRAR) is a Joint Venture of Curtin University and The University of Western Australia, funded by the Western Australian State government. This scientific work makes use of the Murchison Radio-astronomy Observatory, operated by CSIRO. We acknowledge the Wajarri Yamatji people as the traditional owners of the Observatory site. Support for the operation of the MWA is provided by the Australian Government (NCRIS), under a contract to Curtin University administered by Astronomy Australia Limited. We acknowledge the Pawsey Supercomputing Centre, which is supported by the Western Australian and Australian Government.
\end{acknowledgements}

\bibliographystyle{pasa-mnras}
\bibliography{bibli}

\end{document}